\documentclass[12pt,thmsa]{article}
\usepackage{amssymb}
\usepackage{epsfig}

\usepackage{sw20jart}

\input{tcilatex}
\begin{document}

\title{Gravitational Waveforms from the Evaporating ACO Cosmic String Loop}
\author{Malcolm Anderson \\
Department of Mathematics\\
Universiti Brunei Darussalam\\
Jalan Tungku Link, Gadong BE 1410\\
Negara Brunei Darussalam}
\date{}
\maketitle

\begin{abstract}
\noindent The linearly polarized gravitational waveforms from a certain type
of rotating, evaporating cosmic string -- the Allen-Casper-Ottewill loop --
are constructed and plotted over the lifetime of the loop. The formulas for
the waveforms are simple and exact, and describe waves which attenuate
self-similarly, with the amplitude and period of the waves falling off
linearly with time.\bigskip

Short Title: Waveforms from a String Loop\bigskip

\noindent PACS\ numbers: 04.25.Nx, 98.80.Cq
\end{abstract}

\section{Introduction}

Cosmic strings are thin filaments of topologically-trapped Higgs field
energy which may have formed at a symmetry-breaking phase transition in the
early Universe (see \cite{Vile-Shell} for a review). In the zero-thickness
approximation they are effectively line singularities whose dynamics and
stress-energy content are governed by the Nambu-Goto action. Knowledge of
the stress-energy tensor can in turn be used to model the string's
gravitational field, and in particular the effects of gravitational
back-reaction on the string's own motion.

In two previous papers \cite{Anderson3, Anderson5} I examined the
back-reaction problem for a certain type of cosmic string loop, the
Allen-Casper-Ottewill (ACO) solution \cite{ACO1}, which in the absence of
gravity is rigidly rotating, and is thought to have the lowest gravitational
radiative efficiency -- and therefore the longest lifetime -- of any
flat-space loop configuration. In these papers, I constructed a solution of
the linearized Einstein equations describing the self-similar evaporation of
the ACO loop, which radiates energy and angular momentum at a constant rate
and evaporates completely in a finite time. The solution can also be matched
to a relic vacuum spacetime across the future light cone of the final
evaporation point of the loop.

Given the complex nature of the Einstein equations for an extended object
like a cosmic string, it is unlikely that a fully non-linear solution to the
back-reaction problem will ever be developed in even the simplest cases, but
fortunately it can be shown that the self-gravity of a GUT string would
almost everywhere be small enough to justify a weak-field treatment \cite
{Quash}. There are thus good reasons for believing that the evaporating ACO
loop described in \cite{Anderson5} very closely approximates the evolution
of a realistic loop configuration, if indeed cosmic strings ever existed in
the early Universe.

In the present paper I continue the analysis of the evaporating ACO loop by
calculating and plotting the linearly polarized waveforms emitted by this
loop. Since the waveforms emitted by an object are discernible only in the
object's wave zone (at distances large compared to characteristic scales
within the object), all the information needed to construct them can be
extracted from the weak-field limit. The resulting analytic expressions for
the waveforms from the evaporating ACO loop -- which are derived in detail
below -- are unexpectedly simple, and are evidence once again of the
tractability of the solution.

Allen and Ottewill in an earlier study of gravitational wave spectra from
string loops \cite{Allen-Ott} have calculated and plotted the waveforms
emitted by the stationary ACO loop, but the expressions they derived are --
unlike the formulas developed below -- infinite Fourier sums, and of course
these omit the amplitude attenuation and frequency acceleration that
accompany the evaporation. Although current limits on the mass per unit of
cosmic strings make it very unlikely that spectra from individual loops will
ever be observed, the calculations that follow offer an unusual example of a
complete analytic set of waveforms from a realistic, extended,
self-gravitating system.

\section{The Evaporating ACO\ Loop}

The world sheet of a zero-thickness cosmic string is the two-dimensional
surface it traces out as it moves, and is described parametrically by a set
of equations of the form $x^{a}=X^{a}(\zeta ^{A})$, where $x^{a}\equiv
[x^{0},x^{1},x^{2},x^{3}]=[t,x,y,z]$ are local coordinates on the
four-dimensional background spacetime $(\mathbf{M},g_{ab})$, and the
parameters $\zeta ^{A}\equiv (\zeta ^{0},\zeta ^{1})$ are gauge coordinates.
In terms of the intrinsic two-metric 
\begin{equation}
\gamma _{AB}=g_{ab}\,X^{a},_{A}\,X^{b},_{B}
\end{equation}
(with $X^{a},_{A}$ shorthand for $\partial X^{a}/\partial \zeta ^{A}$) the
Nambu-Goto action \cite{Nambu, Goto} reads 
\begin{equation}
I=-\mu \int \gamma ^{1/2}d^{2}\zeta \text{,}
\end{equation}
where $\gamma $ denotes $|\det (\gamma _{AB})|$ and $\mu $ is the (constant)
mass per unit length of the string.

The corresponding stress-energy tensor $T^{ab}$ can be constructed from the
first variation of $I$ with respect to $g_{ab}$ in the standard way, and has
the form 
\begin{equation}
T^{ab}(x^{c})=\mu g^{-1/2}\int \gamma ^{1/2}p^{ab}\,\delta
^{(4)}(x^{c}-X^{c})\,d^{2}\zeta \text{,}  \label{stress}
\end{equation}
where $g\equiv |\det (g_{ab})|$ and $p^{ab}\equiv \gamma
^{AB}X^{a},_{A}X^{b},_{B}$ is the projection tensor onto the tangent space
of the world sheet. The equation of motion of the string is similarly
constructed by setting the first variation of $I$ with respect to $X^{a}$ to
zero, and reads 
\begin{equation}
q_{d}^{c}(\gamma ^{CD}X^{d},_{CD}+p^{mn}\Gamma _{mn}^{d})=0\text{,}
\label{eqmo}
\end{equation}
where $q^{ab}\equiv g^{ab}-p^{ab}$ is the orthogonal complement of $p^{ab}$,
and $\Gamma _{bc}^{a}$ is the Christoffel symbol associated with $g_{ab}$.

If the background spacetime is flat, so that $g_{ab}=\eta _{ab}$ where $\eta
_{ab}$ is the Minkowski metric tensor (here taken to have its rigid form $%
\eta _{ab}=\,$diag$\,(1,-1,-1,-1)$ ), then the equation of motion (\ref{eqmo}%
) reduces to the wave equation 
\begin{equation}
X^{a},_{\tau \tau }=X^{a},_{\sigma \sigma }  \label{eqmo-flat}
\end{equation}
provided that the coordinates $(\tau ,\sigma )=(\zeta ^{0},\zeta ^{1})$ are
chosen to satisfy the gauge conditions $X_{\tau }^{2}+X_{\sigma }^{2}=0$ and 
$X_{\tau }\cdot X_{\sigma }=0$ (which is equivalent to choosing $\gamma
_{AB} $ to be diagonal and trace-free).

A particular solution of the flat-space equation of motion (\ref{eqmo-flat})
is the Allen-Casper-Ottewill (ACO) loop \cite{ACO1}, which is described by
the trajectory 
\begin{equation}
X^{a}=\tfrac{1}{4\pi }L_{0}[u+v,\cos v,\sin v,|u|-\tfrac{1}{2}\pi ]^{a}\text{%
,}  \label{ACO-stat}
\end{equation}
where $u=2\pi (\tau +\sigma )/L_{0}$ and $v=2\pi (\tau -\sigma )/L_{0}$, the
constant $L_{0}$ is the invariant length of the loop, and the parameter $%
\tau $ has been identified with the coordinate time $t$. The gauge
coordinate $u$ covers the range $(-\pi ,\pi ]$, with the limit $u\rightarrow
-\pi ^{+}$ coinciding with $u=\pi $, while the range of $v$ is unrestricted.
The ACO loop rotates rigidly with fundamental oscillation period $%
t_{p}=L_{0}/2$, as shown in Figure 1, and has total energy $E=\mu L_{0}$ and
angular momentum $\mathbf{J}=\tfrac{1}{8\pi }\mu L_{0}^{2}\mathbf{\hat{z}}$.
The points at the extreme top and bottom of the loop, which correspond to $%
u=\pi $ and $u=0$ respectively, are known as kinks. Here the tangent vector $%
X_{u}^{a}$ is discontinuous, although the trajectory itself remains
continuous.

Given any flat-space solution of the equation of motion (\ref{eqmo-flat}),
it is possible to calculate the gravitational power radiated by the loop by
taking the Fourier transform $\bar{T}^{ab}$ of the stress-energy tensor (\ref
{stress}) and substituting it into the quadrupole formula for the power per
unit solid angle at future null infinity: 
\begin{equation}
\frac{dP}{d\Omega }=\frac{\omega ^{2}}{\pi }\sum_{m=1}^{\infty }m^{2}[\bar{T}%
^{ab}\bar{T}_{ab}^{*}\mathbf{-}\tfrac{1}{2}|\bar{T}_{a}^{a}\mathbf{|}^{2}]
\end{equation}
where $\omega =2\pi /t_{p}$ is the circular frequency of the source and $m$
is the Fourier wave number. Because $T^{ab}$ is proportional to the string's
mass per unit length $\mu $, the total power $P$ of the loop scales as $\mu
^{2}$, and a dimensionless measure of the power is the loop's radiative
efficiency $\gamma ^{0}=P/\mu ^{2}$. The significance of the ACO loop lies
in the fact that its radiative efficiency $\gamma ^{0}\approx 39.0025$ --
which was first calculated by Allen, Casper and Ottewill \cite{ACO1} in 1994
-- is the lowest of any known cosmic string loop. All other known analytic
and numerically-generated solutions have higher efficiencies \cite{ACO1,
All-Shell, All-Casp}, although a definitive proof that the ACO loop has the
lowest possible value of $\gamma ^{0}$ has not yet been found.

The stress-energy tensor (\ref{stress}) can also be used to estimate the
gravitational field induced by any flat-space loop by inserting $T^{ab}$
into the linearized Einstein equation. In the linearized approximation, the
metric tensor is decomposed in the form $g_{ab}=\eta _{ab}+h_{ab}$, where
the components of $h_{ab}$ are assumed to be small. If $h_{ab}$ is
constrained to satisfy the harmonic gauge conditions $h_{a}^{b},_{b}=\frac{1%
}{2}h,_{a}$ (with $h\equiv h_{b}^{b}$) then at first order in $h_{ab}$ the
Einstein equation $G^{ab}=-8\pi T^{ab}$ reads $\square h_{ab}=-16\pi S_{ab}$%
, where $S_{ab}=T_{ab}-\frac{1}{2}\eta _{ab}T_{c}^{c}$ and $\square \equiv
\partial _{t}^{2}-\nabla ^{2}$ is the flat-space d'Alembertian. The
corresponding retarded solution for $h_{ab}$ is: 
\begin{equation}
h_{ab}(t,\mathbf{x})=-4\int \frac{S_{ab}(t^{\prime },\mathbf{x}^{\prime })}{|%
\mathbf{x}-\mathbf{x}^{\prime }|}\,d^{3}x^{\prime }  \label{h_ab}
\end{equation}
with $x^{a}\equiv [t,\mathbf{x}]^{a}$ and $t^{\prime }=t-|\mathbf{x}-\mathbf{%
x}^{\prime }|$ the retarded time at the source point $\mathbf{x}^{\prime }$.
In view of the equation (\ref{stress}) for $T^{ab}$, the components of $%
h_{ab}$ are all of order $\mu $.

Once $h_{ab}$ has been calculated, it is in principle possible to determine
the gravitational self-force on the loop, and then use the linearized
version of the equation of motion (\ref{eqmo}) to model the effect of the
back-reaction averaged over a single oscillation period of the loop. This
has been done for the ACO loop in \cite{Anderson3}, where it was found that
after a single period $\Delta \tau =t_{p}$ the loop retains its original
shape, but its length $L_{0}$ shrinks by an amount $\Delta L=-\frac{1}{2}%
\gamma ^{0}\mu L_{0}\approx -19.501\mu L_{0}$ while its rotational phase
advances by an angle of about $38.92\mu $ radians. In particular, the
radiative efficiency $\gamma ^{0}$ remains unchanged, and so the ACO\ loop
radiates energy at a constant rate even in the presence of gravitational
back-reaction.

In \cite{Anderson5} I used a simple geometric argument to demonstrate that
if the ACO loop does evaporate self-similarly with a constant power $%
P=\gamma ^{0}\mu ^{2}$ then its trajectory, to leading order in $\mu $, will
have the form 
\begin{equation}
X^{a}(u,v)=[t_{L},\mathbf{0}]^{a}+\tfrac{1}{4\pi }L_{0}\,e^{-\kappa \mu
(u+v)}[-\tfrac{1}{\kappa \mu },\cos v,\sin v,|u|-\tfrac{1}{2}\pi ]^{a}
\label{ACO-evap}
\end{equation}
where $\kappa \equiv \gamma ^{0}/(4\pi )$ and $t_{L}\equiv \tfrac{1}{4\pi
\kappa \mu }L_{0}$ is the total lifetime of a loop whose length at time $t=0$
is $L_{0}$. As in the case of the stationary loop (\ref{ACO-stat}), the
coordinate $u$ covers the range $(-\pi ,\pi ]$. The evaporating loop (\ref
{ACO-evap}) shrinks to a point at $x^{a}=[t_{L},\mathbf{0}]^{a}$ in the
limit as $v\rightarrow \infty $. Furthermore, in the neighborhood of any
spacelike surface defined by $t=t_{L}(1-e^{-\kappa \mu v_{*}})$, where $%
v_{*} $ is an arbitrary constant, the evaporating trajectory (\ref{ACO-evap}%
) is isometric to the stationary trajectory (\ref{ACO-stat}) to leading
order in $\mu $, save with $L_{0}$ replaced by $L_{*}=L_{0}e^{-\kappa \mu
v_{*}}$, and $\cos v$ and $\sin v$ replaced by cos$(v_{*}+v)$ and sin$%
(v_{*}+v)$. The evaporating trajectory therefore satisfies the flat-space
equation of motion (\ref{eqmo-flat}) to leading order in $\mu $.

As with the stationary ACO loop, it is possible to use the retarded integral
(\ref{h_ab}) to calculate the metric perturbations $h_{ab}$ induced by the
evaporating loop (\ref{ACO-evap}). The exact form of these perturbations
will be presented in Section 3. The perturbations were used to develop the
string equation of motion (\ref{eqmo}) to order $\mu $ in \cite{Anderson5},
where after a lengthy calculation it was shown that the trajectory of the
evaporating ACO loop, correct now to order $\mu ^{2}$, has the form 
\begin{equation}
X^{a}(u,v)=[t_{L},\mathbf{0}]^{a}+\tfrac{1}{4\pi }L_{0}\,e^{-\kappa \mu
(u+v)}[-\tfrac{1}{\kappa \mu }e^{\kappa \mu ^{2}F(u)},\mathbf{\bar{r}}%
(u,v)]^{a}  \label{ACO-2}
\end{equation}
with 
\begin{equation}
\mathbf{\bar{r}}(u,v)=e^{\mu G(u)}[\cos (v)\,\mathbf{\hat{x}}+\sin (v)\,%
\mathbf{\hat{y}]}+[(|u|-\tfrac{1}{2}\pi )+\mu A(u)]\,\mathbf{\hat{z}}\text{,}
\end{equation}
where the functions $F$, $G$ (both periodic in $u$ with period $\pi $) and $%
A $ (anti-periodic in $u$, with $A(0)=-A(\pi )=-12\pi +8\pi \ln 2\pi +2\func{%
Si}(2\pi )$) are fixed up to a Poincar\'{e} transformation and an arbitrary
gauge transformation of $A$. In addition, the expected value of $\kappa
\equiv \gamma ^{0}/(4\pi )$ -- namely $4[\ln 2\pi +\gamma _{\text{E}}-\func{%
Ci}(2\pi )]/\pi \approx 3.\,103\,7$ (where $\gamma _{\text{E}}$ is Euler's
constant) -- emerges as a consistency condition on the solution (\ref{ACO-2}%
).

It was further demonstrated in \cite{Anderson5} that the metric
perturbations $h_{ab}$ can be matched smoothly (to order $\mu $) to a
Minkowski spacetime across the future light cone $F_{L}=\{t-t_{L}=|\mathbf{x}%
|\}$ of the final evaporation point at $x^{a}=[t_{L},\mathbf{0}]^{a}$, and
so the evolving ACO loop (\ref{ACO-evap}) evaporates completely to leave
behind a remnant vacuum spacetime, as would be expected. The gross features
of the loop's evaporation are illustrated in Figure 2.

\section{The Far-Field Metric Perturbations}

If $[t,\mathbf{x}]$ is a general field point satisfying the constraint $%
t-t_{L}<|\mathbf{x}|$ (and so lying to the past of the null surface $F_{L}$%
), then the linearized metric perturbations $h_{ab}$ due to the evaporating
ACO loop (\ref{ACO-evap}) have the self-similar form \cite{Anderson5} 
\begin{equation}
h_{ab}(t,\mathbf{x})=4\mu \left[ 
\begin{array}{cccc}
I_{+}+I_{-} & \bar{S}_{+}+\bar{S}_{-} & -\bar{C}_{+}-\bar{C}_{-} & 
I_{+}-I_{-} \\ 
\bar{S}_{+}+\bar{S}_{-} & I_{+}+I_{-} & 0 & \bar{S}_{+}-\bar{S}_{-} \\ 
-\bar{C}_{+}-\bar{C}_{-} & 0 & I_{+}+I_{-} & -\bar{C}_{+}+\bar{C}_{-} \\ 
I_{+}-I_{-} & \bar{S}_{+}-\bar{S}_{-} & -\bar{C}_{+}+\bar{C}_{-} & 
I_{+}+I_{-}
\end{array}
\right] _{ab}  \label{h_ab-new}
\end{equation}
where $I_{\pm }$, $\bar{S}_{\pm }$ and $\bar{C}_{\pm }$ are functions of the
rotation angle 
\begin{equation}
\psi (t,\mathbf{x})\equiv -(\kappa \mu )^{-1}\ln [(|\mathbf{x}%
|-t+t_{L})/t_{L}]  \label{psi}
\end{equation}
and the components of an image point $\bar{x}^{a}=[\bar{t},\mathbf{\bar{x}}%
]^{a}$ lying on the future light cone $F_{0}=\{t=|\mathbf{x}|\}$ of the
origin, and defined by 
\[
\bar{t}=[(|\mathbf{x}|-t+t_{L})/t_{L}]^{-1}|\mathbf{x}|\text{,\qquad }\bar{x}%
=[(|\mathbf{x}|-t+t_{L})/t_{L}]^{-1}(x\cos \psi +y\sin \psi ) 
\]
\begin{equation}
\bar{y}=[(|\mathbf{x}|-t+t_{L})/t_{L}]^{-1}(y\cos \psi -x\sin \psi )\qquad 
\text{and\qquad }\bar{z}=[(|\mathbf{x}|-t+t_{L})/t_{L}]^{-1}z\text{.}
\label{barred}
\end{equation}

To be specific, if $[T,X,Y,Z]\equiv 4\pi L_{0}^{-1}[\bar{t},\bar{x},\bar{y},%
\bar{z}]$ and the functions $V_{0}$ and $V_{1}$ are defined implicitly by
the equations 
\begin{equation}
T-V_{0}=[(X-\cos V_{0})^{2}+(Y-\sin V_{0})^{2}+(Z+\tfrac{1}{2}\pi
)^{2}]^{1/2}  \label{V_0}
\end{equation}
and 
\begin{equation}
T-(\pi +V_{1})=[(X-\cos V_{1})^{2}+(Y-\sin V_{1})^{2}+(Z-\tfrac{1}{2}\pi
)^{2}]^{1/2}\text{,}  \label{V_1}
\end{equation}
then $I_{\pm }$, $\bar{S}_{\pm }$ and $\bar{C}_{\pm }$ have the explicit
forms 
\begin{equation}
I_{+}=\ln [(\chi _{+}-V_{1}-2\pi )/(\chi _{+}-V_{0})]\text{,\qquad }%
I_{-}=\ln [(\chi _{-}-V_{0})/(\chi _{-}-V_{1})]\text{,}  \label{I}
\end{equation}
\begin{equation}
\left[ 
\begin{array}{c}
\bar{S}_{+} \\ 
\bar{C}_{+}
\end{array}
\right] =\left[ 
\begin{array}{cc}
-\cos (\chi _{+}+\psi ) & \sin (\chi _{+}+\psi ) \\ 
\sin (\chi _{+}+\psi ) & \cos (\chi _{+}+\psi )
\end{array}
\right] \left[ 
\begin{array}{c}
\func{Si}(\chi _{+}-V_{1}-2\pi )-\func{Si}(\chi _{+}-V_{0}) \\ 
\func{Ci}(\chi _{+}-V_{1}-2\pi )-\func{Ci}(\chi _{+}-V_{0})
\end{array}
\right]  \label{plus}
\end{equation}
and 
\begin{equation}
\left[ 
\begin{array}{c}
\bar{S}_{-} \\ 
\bar{C}_{-}
\end{array}
\right] =\left[ 
\begin{array}{cc}
-\cos (\chi _{-}+\psi ) & \sin (\chi _{-}+\psi ) \\ 
\sin (\chi _{-}+\psi ) & \cos (\chi _{-}+\psi )
\end{array}
\right] \left[ 
\begin{array}{c}
\func{Si}(\chi _{-}-V_{0})-\func{Si}(\chi _{-}-V_{1}) \\ 
\func{Ci}(\chi _{-}-V_{0})-\func{Ci}(\chi _{-}-V_{1})
\end{array}
\right] \text{,}  \label{minus}
\end{equation}
where $\chi _{\pm }\equiv T\pm (Z+\tfrac{\pi }{2})$, and $\func{Si}%
(x)=\int_{0}^{x}w^{-1}\sin w\,dw$ and $\func{Ci}(x)=-\int_{x}^{\infty
}w^{-1}\cos w\,dw$ are the usual sine and cosine integrals. Note also that $%
T=(X^{2}+Y^{2}+Z^{2})^{1/2}$, as $\bar{x}^{a}$ lies on the light cone $F_{0}$%
.

The bars on $\bar{S}_{\pm }$ and $\bar{C}_{\pm }$ have here been used to
distinguish them from the corresponding unbarred quantities introduced in 
\cite{Anderson5}, which were defined analogously to (\ref{plus}) and (\ref
{minus}) except with $\psi =0$. Apart from this explicit $\psi $ dependence,
the metric perturbations defined by (\ref{h_ab-new}) are determined
completely by the components of the image point $\bar{x}^{a}$ alone. It is
for this reason that the weak-field metric was described in \cite{Anderson5}
as exhibiting ``rotating self-similarity''.\footnote{%
In fact, it can be shown that the vector field 
\[
k^{a}=[0,-y,x,0]^{a}-\kappa \mu [t-t_{L},x,y,z]^{a}\text{,} 
\]
which is constructed by taking $k^{a}=\partial x^{a}/\partial \psi $ (with
the components of the image point $\bar{x}^{a}$ kept fixed), is a \emph{%
conformal} Killing vector field of the linearized metric $g_{ab}=\eta
_{ab}+h_{ab}$, as it satisfies the equation 
\[
k_{(a;b)}=-\kappa \mu g_{ab} 
\]
to linear order in $\mu $.}

The main purpose of this paper is to calculate the gravitational waveforms
generated by the metric perturbations $h_{ab}$ as measured at a field point
at a fixed distance $r\equiv |\mathbf{x}|$ from $\mathbf{x}=\mathbf{0}$. It
will henceforth be assumed that the field point lies in the wave zone, where 
$r\gg L_{0}$. To calculate the waveforms it is necessary to know the metric
components only to linear order in $\mu $, and these only to order $r^{-1}$.

It should be noted, however, that the quantity $r-t$ appearing in the
definition (\ref{psi}) of $\psi $ and subsequent equations is not
necessarily large. In fact, the combination $(t-r)/t_{L}$ is a convenient
measure of observer time at the field point, and ranges from $0$ on the
initial null surface $F_{0}$ to $1$ in the limit as the field point
approaches the boundary null surface $F_{L}$.

In view of (\ref{barred}), therefore, the components of the image point $%
\bar{x}^{a}$ are all of order $r$ in the wave zone, and the equations (\ref
{V_0}) and (\ref{V_1}) defining $V_{0}$ and $V_{1}$ reduce to 
\begin{equation}
V_{0}=r^{-1}[x\cos (V_{0}+\psi )+y\sin (V_{0}+\psi )-\tfrac{\pi }{2}z]
\label{asympV_0}
\end{equation}
and 
\begin{equation}
V_{1}+\pi =r^{-1}[x\cos (V_{1}+\psi )+y\sin (V_{1}+\psi )+\tfrac{\pi }{2}z]
\label{asympV_1}
\end{equation}
respectively. Although the functions $V_{0}$ and $V_{1}$ typically do not
converge as the field point approaches $F_{L}$ (where $\psi \rightarrow
\infty $), it is clear that $|V_{0}|$ and $|V_{1}+\pi |$ are both bounded
above by $(1+\frac{1}{4}\pi ^{2})^{1/2}$.

As the asymptotic behavior of the metric perturbations $h_{ab}$ is dominated
by the quantities $\chi _{+}$ and $\chi _{-}$, the cases in which the field
point does or does not lie on the $z$-axis (where $x=y=0$ and so $T=|Z|$)
need to be considered separately.

If the field point does lie on the $z$-axis, then (\ref{asympV_0})\ and (\ref
{asympV_1}) immediately reduce to $V_{0}=-\tfrac{\pi }{2}\xi $ and $V_{1}=%
\tfrac{\pi }{2}\xi -\pi $, where $\xi \equiv \,$sgn$(z)$, while $\chi _{\pm
}=\xi Z\pm (Z+\tfrac{\pi }{2})$. However, two of the metric functions $\bar{S%
}_{\pm }$ and $\bar{C}_{\pm }$ are then undefined, which is simply an
indication that the asymptotic expansions (\ref{asympV_0}) and (\ref
{asympV_1}) are too coarse.

Fortunately, it is possible to solve equations (\ref{V_0}) and (\ref{V_1})
explicitly in this case to give \cite{Anderson5} 
\begin{equation}
V_{0}=\xi Z-[1+(Z+\tfrac{\pi }{2})^{2}]^{1/2}\qquad \text{and}\qquad
V_{1}=\xi Z-\pi -[1+(Z-\tfrac{\pi }{2})^{2}]^{1/2}\text{.}
\end{equation}
The metric perturbation $h_{tz}$ is then identically zero, while the
remaining, non-zero components have the asymptotic forms 
\begin{equation}
h_{tt}=h_{xx}=h_{yy}=h_{zz}=-2\mu L/r+O(r^{-2})\text{,}
\end{equation}
\begin{equation}
\xi h_{tx}=-h_{zx}=(\cos \psi )\mu L/r+O(r^{-2})
\end{equation}
and 
\begin{equation}
\xi h_{ty}=-h_{zy}=(\sin \psi )\mu L/r+O(r^{-2})\text{,}
\end{equation}
where $L$ is the parametric length of the string at the time of emission of
the waveform: 
\begin{equation}
L(t)=L_{0}[(r-t+t_{L})/t_{L}]\text{,}
\end{equation}
In fact, as will be seen in the next section, the corresponding waveforms
carry no gravitational energy along the $z$-axis.

In the more general case in which the field point lies off the $z$-axis, the
functions $I_{\pm }$, $\bar{S}_{\pm }$ and $\bar{C}_{\pm }$ can be
represented by the leading-order expressions 
\begin{equation}
I_{+}=(1+\cos \theta )^{-1}(V_{0}-V_{1}-2\pi )L/(4\pi r)\text{,}
\end{equation}
\begin{equation}
I_{-}=(1-\cos \theta )^{-1}(V_{0}-V_{1})L/(4\pi r)\text{,}
\end{equation}
\begin{equation}
\left[ 
\begin{array}{c}
\bar{S}_{+} \\ 
\bar{C}_{+}
\end{array}
\right] =\left[ 
\begin{array}{c}
\cos (V_{1}+\psi )-\cos (V_{0}+\psi ) \\ 
-\sin (V_{1}+\psi )+\sin (V_{0}+\psi )
\end{array}
\right] (1+\cos \theta )^{-1}L/(4\pi r)
\end{equation}
and 
\begin{equation}
\left[ 
\begin{array}{c}
\bar{S}_{-} \\ 
\bar{C}_{-}
\end{array}
\right] =\left[ 
\begin{array}{c}
-\cos (V_{1}+\psi )+\cos (V_{0}+\psi ) \\ 
\sin (V_{1}+\psi )-\sin (V_{0}+\psi )
\end{array}
\right] (1-\cos \theta )^{-1}L/(4\pi r)\text{,}
\end{equation}
where $\cos \theta =z/r$.

The components of the $h_{ab}$ therefore have the asymptotic forms: 
\begin{equation}
h_{tt}=h_{xx}=h_{yy}=h_{zz}=\frac{2\mu L}{\pi r\sin ^{2}\theta }[-\pi
+(V_{1}+\pi -V_{0})\cos \theta ]\text{,}
\end{equation}
\begin{equation}
h_{tz}=\frac{2\mu L}{\pi r\sin ^{2}\theta }(V_{0}-V_{1}-\pi +\pi \cos \theta
)\text{,}  \label{h_tz}
\end{equation}
\begin{equation}
h_{tx}=\frac{2\mu L\cos \theta }{\pi r\sin ^{2}\theta }[-\cos (V_{1}+\psi
)+\cos (V_{0}+\psi )]\text{,}
\end{equation}
\begin{equation}
h_{ty}=\frac{2\mu L\cos \theta }{\pi r\sin ^{2}\theta }[-\sin (V_{1}+\psi
)+\sin (V_{0}+\psi )]\text{,}
\end{equation}
\begin{equation}
h_{zx}=\frac{2\mu L}{\pi r\sin ^{2}\theta }[\cos (V_{1}+\psi )-\cos
(V_{0}+\psi )]
\end{equation}
and 
\begin{equation}
h_{zy}=\frac{2\mu L}{\pi r\sin ^{2}\theta }[\sin (V_{1}+\psi )-\sin
(V_{0}+\psi )]\text{.}  \label{h_zy}
\end{equation}

\section{Calculating the Waveforms}

The procedure for calculating the linearly polarized gravitational waveforms 
$h_{+}$ and $h_{\times }$ from a knowledge of $h_{ab}$ has been conveniently
summarized by Allen and Ottewill in \cite{Allen-Ott}. This reference
generates the waveforms for a number of simple cosmic string loop
configurations, including the stationary ACO loop (\ref{ACO-stat}). However,
the waveforms for the ACO\ loop are there presented in the form of series
approximations, as the metric perturbations $h_{ab}$ are calculated by means
of Fourier expansions. By contrast, the waveforms constructed in this
section are those from the \emph{evaporating} ACO loop (\ref{ACO-evap}), and
these can be cast in a relatively simple closed form, as will be seen
shortly.

The first step in calculating the waveforms is to construct the radiative
perturbations $\bar{h}_{ab}$ formed by adding the potential term $2\mu L/r$
to each of the diagonal elements of $h_{ab}$, so that 
\begin{equation}
\bar{h}_{ab}=h_{ab}+(2\mu L/r)\delta _{ab}\text{.}
\end{equation}

In the case where the field point lies on the $z$-axis this yields the very
simple radiative metric 
\begin{equation}
\bar{h}_{ab}=(\mu L/r)\left[ 
\begin{array}{cccc}
0 & \xi \cos \psi & \xi \sin \psi & 0 \\ 
\xi \cos \psi & 0 & 0 & -\cos \psi \\ 
\xi \sin \psi & 0 & 0 & -\sin \psi \\ 
0 & -\cos \psi & -\sin \psi & 0
\end{array}
\right] \text{.}
\end{equation}
On the other hand, if the field point does not lie on the $z$-axis then 
\begin{equation}
\bar{h}_{tt}=\bar{h}_{xx}=\bar{h}_{yy}=\bar{h}_{zz}=\frac{2\mu L\cos \theta 
}{\pi r\sin ^{2}\theta }(V_{1}+\pi -V_{0}-\pi \cos \theta )\text{,}
\end{equation}
whereas the remaining, off-diagonal components (\ref{h_tz})-(\ref{h_zy}) are
unchanged: 
\begin{equation}
\bar{h}_{ab}=h_{ab}\qquad \text{if\quad }a\neq b\text{.}
\end{equation}

The next step is to eliminate the time-time and time-space components of $%
\bar{h}_{ab}$ by performing the gauge transformation 
\begin{equation}
\bar{h}_{ab}^{\prime }=\bar{h}_{ab}-n_{a}\varepsilon _{b}-n_{b}\varepsilon
_{a}\text{,}  \label{hprime}
\end{equation}
where $n^{a}=[1,\mathbf{n}]^{a}$ is a null vector in the direction of the
field point $x^{a}$, and $\varepsilon _{a}=[\varepsilon _{0},\mathbf{e}]_{a}$
with 
\begin{equation}
\varepsilon _{0}=\tfrac{1}{2}\bar{h}_{tt}
\end{equation}
and the components of the 3-vector $\mathbf{e}$ given by 
\begin{equation}
e_{j}=-\tfrac{1}{2}\bar{h}_{tt}n_{j}+\bar{h}_{tj}\text{.}
\end{equation}

If the field point is represented in standard spherical polar coordinates as 
\begin{equation}
x^{a}=[t,r\sin \theta \cos \phi ,r\sin \theta \sin \phi ,r\cos \theta ]^{a}
\end{equation}
then 
\begin{equation}
\mathbf{n}=(\sin \theta \cos \phi ,\sin \theta \sin \phi ,\cos \theta )
\end{equation}
(and the quantities $n_{j}$ are the components of $-\mathbf{n}$).

In the case where the field point lies on the $z$-axis, $\varepsilon _{0}=0$
and $\mathbf{n}=(0,0,\xi )$, while 
\begin{equation}
e_{j}=(\mu L/r)(\xi \cos \psi ,\xi \sin \psi ,0)_{j}\text{.}
\end{equation}
It is easily verified that in this case $\bar{h}_{ab}^{\prime }=0$ and so
the metric perturbations are pure gauge terms. An immediate consequence is
that the evaporating ACO loop radiates no gravitational energy along the $z$%
-axis, as mentioned earlier.

In the more general case where the field point lies off the $z$-axis, it
proves convenient to eliminate the term $V_{1}+\pi -V_{0}$ appearing in $%
\bar{h}_{tt}$ and $\bar{h}_{tz}$ by invoking equations (\ref{asympV_0}) and (%
\ref{asympV_1}), which can be written more compactly as 
\begin{equation}
V_{0}=\cos (V_{0}+\psi -\phi )\sin \theta -\tfrac{\pi }{2}\cos \theta
\label{newV_0}
\end{equation}
and 
\begin{equation}
V_{1}+\pi =\cos (V_{1}+\psi -\phi )\sin \theta +\tfrac{\pi }{2}\cos \theta 
\text{.}  \label{newV_1}
\end{equation}

Then 
\begin{equation}
\bar{h}_{tt}=\bar{h}_{xx}=\bar{h}_{yy}=\bar{h}_{zz}=\frac{2\mu L\cos \theta 
}{\pi r\sin \theta }(\kappa _{1}-\kappa _{0})
\end{equation}
and 
\begin{equation}
\bar{h}_{tz}=-\frac{2\mu L}{\pi r\sin \theta }(\kappa _{1}-\kappa _{0})\text{%
,}
\end{equation}
where 
\begin{equation}
\kappa _{0},_{1}\equiv \cos (V_{0},_{1}+\psi -\phi )\text{.}
\end{equation}

Hence, 
\begin{eqnarray}
\varepsilon _{a} &=&\frac{\mu L}{\pi r\sin ^{2}\theta }[(\kappa _{1}-\kappa
_{0})\sin \theta \cos \theta ,\{(\kappa _{1}-\kappa _{0})\sin ^{2}\theta
\cos \phi -2(c_{1}-c_{0})\}\cos \theta ,  \nonumber \\
&&\{(\kappa _{1}-\kappa _{0})\sin ^{2}\theta \sin \phi -2(s_{1}-s_{0})\}\cos
\theta ,-(\kappa _{1}-\kappa _{0})(2-\cos ^{2}\theta )\sin \theta ]_{a}\text{%
,}  \nonumber \\
&&
\end{eqnarray}
with 
\begin{equation}
c_{0},_{1}\equiv \cos (V_{0},_{1}+\psi )\qquad \text{and}\qquad
s_{0},_{1}\equiv \sin (V_{0},_{1}+\psi )\text{.}
\end{equation}

The space-space components of the corresponding transformed metric
perturbations $\bar{h}_{ab}^{\prime }$ are: 
\begin{equation}
\bar{h}_{xx}^{\prime }=\frac{2\mu L\cos \theta }{\pi r\sin \theta }[(\kappa
_{1}-\kappa _{0})(1+\sin ^{2}\theta \cos ^{2}\phi )-2(c_{1}-c_{0})\cos \phi ]%
\text{,}  \label{hprime-first}
\end{equation}
\begin{equation}
\bar{h}_{yy}^{\prime }=\frac{2\mu L\cos \theta }{\pi r\sin \theta }[(\kappa
_{1}-\kappa _{0})(1+\sin ^{2}\theta \sin ^{2}\phi )-2(s_{1}-s_{0})\sin \phi ]%
\text{,}
\end{equation}
\begin{equation}
\bar{h}_{zz}^{\prime }=-\frac{2\mu L}{\pi r}(\kappa _{1}-\kappa _{0})\sin
\theta \cos \theta \text{,}
\end{equation}
\begin{equation}
\bar{h}_{xy}^{\prime }=\frac{2\mu L\cos \theta }{\pi r\sin \theta }[-(\kappa
_{1}-\kappa _{0})(1+\cos ^{2}\theta )\sin \phi \cos \phi +(1-2\cos ^{2}\phi
)(\sigma _{1}-\sigma _{0})]\text{,}
\end{equation}
\begin{equation}
\bar{h}_{xz}^{\prime }=\frac{2\mu L}{\pi r}[(c_{1}-c_{0})-(\kappa
_{1}-\kappa _{0})\sin ^{2}\theta \cos \phi ]
\end{equation}
and 
\begin{equation}
\bar{h}_{yz}^{\prime }=\frac{2\mu L}{\pi r}[(s_{1}-s_{0})-(\kappa
_{1}-\kappa _{0})\sin ^{2}\theta \sin \phi ]\text{,}  \label{hprime-last}
\end{equation}
where now 
\begin{equation}
\sigma _{0},_{1}\equiv \sin (V_{0},_{1}+\psi -\phi )\text{.}
\end{equation}

An important feature of the 3-tensor $\bar{h}_{jk}^{\prime }$ is that the
contraction $n^{j}\bar{h}_{jk}^{\prime }$ is identically zero, as can be
verified directly from the expressions (\ref{hprime-first})-(\ref
{hprime-last}). The metric perturbations $\bar{h}_{jk}^{\prime }$ can be
reduced to canonical linearly polarized form by rotating the three spatial
axes so that the new $z$-axis is aligned with the unit vector $\mathbf{n}$
along the line of sight. This is accomplished by introducing an orthonormal
triad $\{\mathbf{m}_{1},\mathbf{m}_{2},\mathbf{n}\}$ with 
\begin{equation}
\mathbf{m}_{2}=(-\sin \phi ,\cos \phi ,0)
\end{equation}
and 
\begin{equation}
\mathbf{m}_{1}=\mathbf{m}_{2}\times \mathbf{n}=(\cos \theta \cos \phi ,\cos
\theta \sin \phi ,-\sin \theta )\text{.}
\end{equation}
The original $x$, $y$ and $z$ coordinates can be aligned with the vectors $%
\mathbf{m}_{1}$, $\mathbf{m}_{2}$ and $\mathbf{n}$ by first rotating the
axes counterclockwise by an angle $\phi $ in the $x$-$y$ plane (thus
aligning the new $y$-axis with $\mathbf{m}_{2}$), then rotating
counterclockwise by an angle $\theta $ about $\mathbf{m}_{2}$.

Given that $\bar{h}_{jk}^{\prime }$ is orthogonal to $\mathbf{n}$, the
perturbation tensor can be decomposed as the linear combination 
\begin{equation}
\bar{h}_{jk}^{\prime
}=Pm_{1j}m_{1k}+Q(m_{1j}m_{2k}+m_{2j}m_{1k})+Rm_{2j}m_{2k}\text{,}
\end{equation}
where 
\begin{equation}
P=\bar{h}_{jk}^{\prime }m_{1}^{j}m_{1}^{k}=-\frac{2\mu L\cos \theta }{\pi
r\sin \theta }(\kappa _{1}-\kappa _{0})\text{,}
\end{equation}
\begin{equation}
Q=\bar{h}_{jk}^{\prime }m_{1}^{j}m_{2}^{k}=-\frac{2\mu L}{\pi r\sin \theta }%
(\sigma _{1}-\sigma _{0})
\end{equation}
and 
\begin{equation}
R=\bar{h}_{jk}^{\prime }m_{2}^{j}m_{2}^{k}=\frac{2\mu L\cos \theta }{\pi
r\sin \theta }(\kappa _{1}-\kappa _{0})\text{.}
\end{equation}

Thus $P\equiv -R$ and is conventionally denoted as $h_{+}$, while $Q$ is
usually denoted as $h_{\times }$. The linearly polarized waveforms therefore
have the very simple forms\footnote{%
Note that a much quicker and direct way of calculating $h_{+}$ and $%
h_{\times }$ is to introduce the 4-vectors $m_{1}^{a}=[0,\mathbf{m}_{1}]^{a}$
and $m_{2}^{a}=[0,\mathbf{m}_{2}]^{a}$. Since $m_{1}^{a}n_{a}=$ $%
m_{2}^{a}n_{a}=0$ it follows immediately from (\ref{hprime}) that 
\[
h_{+}=\bar{h}_{ab}m_{1}^{a}m_{1}^{a} 
\]
and 
\[
h_{\times }=\bar{h}_{ab}m_{1}^{a}m_{2}^{a}\text{.} 
\]
} 
\begin{equation}
h_{+}=-\frac{2\mu L_{0}\cos \theta }{\pi r\sin \theta }[(r-t+t_{L})/t_{L}][%
\cos (V_{1}+\psi -\phi )-\cos (V_{0}+\psi -\phi )]  \label{hplus}
\end{equation}
and 
\begin{equation}
h_{\times }=-\frac{2\mu L_{0}}{\pi r\sin \theta }[(r-t+t_{L})/t_{L}][\sin
(V_{1}+\psi -\phi )-\sin (V_{0}+\psi -\phi )]\text{,}  \label{hcross}
\end{equation}
where it should be recalled that $\psi =-(\kappa \mu )^{-1}\ln
[(r-t+t_{L})/t_{L}]$ is a known function of the observer time $t$, while $%
V_{1}$ and $V_{0}$ are implicit functions of $t$, $\theta $ and $\phi $
determined by (\ref{newV_0}) and (\ref{newV_1}).

The expressions (\ref{hplus}) and (\ref{hcross}) for $h_{+}$ and $h_{\times
} $ were of course generated on the assumption that the observer's $y$-axis
(which is $\mathbf{m}_{2}$) is parallel to the $x$-$y$ plane of the loop. In
the more general situation, the observer's $x$- and $y$-axes will be rotated
by some angle $\vartheta $ counterclockwise about $\mathbf{n}$ from the
positions of $\mathbf{m}_{1}$ and $\mathbf{m}_{2}$. In such a case, the
waveform matrix will have the form 
\begin{eqnarray}
&&\left[ 
\begin{array}{ccc}
\cos \vartheta & -\sin \vartheta & 0 \\ 
\sin \vartheta & \cos \vartheta & 0 \\ 
0 & 0 & 1
\end{array}
\right] \left[ 
\begin{array}{ccc}
h_{+} & h_{\times } & 0 \\ 
h_{\times } & -h_{+} & 0 \\ 
0 & 0 & 0
\end{array}
\right] \left[ 
\begin{array}{ccc}
\cos \vartheta & -\sin \vartheta & 0 \\ 
\sin \vartheta & \cos \vartheta & 0 \\ 
0 & 0 & 1
\end{array}
\right] ^{T}  \nonumber \\
&=&\left[ 
\begin{array}{ccc}
h_{+}\cos 2\vartheta -h_{\times }\sin 2\vartheta & h_{\times }\cos
2\vartheta +h_{+}\sin 2\vartheta & 0 \\ 
h_{\times }\cos 2\vartheta +h_{+}\sin 2\vartheta & -h_{+}\cos 2\vartheta
+h_{\times }\sin 2\vartheta & 0 \\ 
0 & 0 & 0
\end{array}
\right] \text{,}
\end{eqnarray}
and so the linearly polarized waveforms are 
\begin{equation}
h_{+}^{\prime }=h_{+}\cos 2\vartheta -h_{\times }\sin 2\vartheta \text{%
\qquad and\qquad }h_{\times }^{\prime }=h_{\times }\cos 2\vartheta
+h_{+}\sin 2\vartheta \text{.}
\end{equation}
The factor of $2$ multiplying the rotation angle $\vartheta $ is just a
reflection of the fact that the graviton is a spin-$2$ particle.

\section{Plotting the Waveforms}

All that remains now is to plot the linearly polarized waveforms $h_{+}$ and 
$h_{\times }$ as functions of $\frak{T}\equiv (t-r)/t_{L}$ for some sample
values of the observer latitude $\theta $ and rotation angle $\vartheta $.
Note from equations (\ref{hplus}) and (\ref{hcross}) that both waveforms are
bounded by an envelope that falls off linearly with $\frak{T}$, but are
otherwise periodic in $\psi =-(\kappa \mu )^{-1}\ln [(r-t+t_{L})/t_{L}]$
with period $2\pi $ (as follows from the fact $V_{0}$ and $V_{1}$ are also
periodic in $\psi $). Since different choices of the observer longitude $%
\phi $ simply alter the phase of the waveforms within the envelope, the
shapes of the waveforms are easily generalized from an examination of the
case $\phi =0$ only.

The most problematic part of the calculation is solving the equations (\ref
{newV_0}) and (\ref{newV_1}) for $V_{0}$ and $V_{1}$. A natural method of
solving equation (\ref{newV_0}) is to use the Newton-Raphson algorithm,
which involves iterating the recurrence relation 
\begin{equation}
v_{n+1}=v_{n}-\frac{v_{n}-\cos (v_{n}+\psi )\sin \theta +\tfrac{\pi }{2}\cos
\theta }{1+\sin (v_{n}+\psi )\sin \theta }\text{,}  \label{Newt-Raph}
\end{equation}
with $v_{0}=-\tfrac{\pi }{2}\cos \theta $. If the sequence $\{v_{n}\}$
converges, it will converge to $V_{0}$. Note from (\ref{newV_0}) and (\ref
{newV_1}) that if $V_{0}$ is known as a function of $\psi $ then $V_{1}(\psi
)=-\pi -V_{0}(-\psi )$ (on the assumption that $\phi =0$).

In the particular case $\theta =\pi /2$ the recurrence relation (\ref
{Newt-Raph}) does not converge, as the denominator is not bounded away from
zero. An alternative is to solve for $V_{0}$ by simple iteration, using the
recurrence relation 
\begin{equation}
v_{n+1}=\cos (v_{n}+\psi )  \label{simple}
\end{equation}
with $v_{0}=0$. This recurrence relation is of course not a contraction
mapping, and it seems not to converge without a little tweaking (entering
instead a limit cycle with period 2). Fortunately, it does seem to converge
if $v_{3}$ is set equal to $\frac{1}{2}(v_{2}+v_{1})$, rather than
calculated from (\ref{simple}), and the algorithm continued as before from
there.

For small values of $\theta $, it is possible to develop power series
solutions of equations (\ref{newV_0}) and (\ref{newV_1}) for $V_{0}$ and $%
V_{1}$, which when $\phi =0$ read: 
\begin{equation}
V_{0}=-\tfrac{\pi }{2}+\theta \sin \psi +\theta ^{2}(\tfrac{1}{4}\pi +\cos
\psi \sin \psi )+O(\theta ^{3})
\end{equation}
and 
\begin{equation}
V_{1}=-\tfrac{\pi }{2}+\theta \sin \psi -\theta ^{2}(\tfrac{1}{4}\pi -\cos
\psi \sin \psi )+O(\theta ^{3})\text{.}
\end{equation}
The corresponding asymptotic expansions for the waveforms $h_{+}$ and $%
h_{\times }$ therefore describe, to leading order in $\theta $, modulated
cosine and sine waves in $\psi $: 
\begin{equation}
h_{+}=\frac{\mu L_{0}\theta }{r}[(r-t+t_{L})/t_{L}]\cos \psi +O(\theta ^{2})
\label{hplus-asymp}
\end{equation}
and 
\begin{equation}
h_{\times }=\frac{\mu L_{0}\theta }{r}[(r-t+t_{L})/t_{L}]\sin \psi +O(\theta
^{2})\text{.}  \label{hcross-asymp}
\end{equation}
Note in particular that the amplitudes of both waveforms tend to zero as $%
\theta \rightarrow 0$, as is to be expected from the fact, mentioned
earlier, that there is no flux of gravitational energy along the $z$-axis.

In Figure 3a to 3d the waveforms $h_{+}$ and $h_{\times }$ (which by
definition have a rotation angle $\vartheta =0$), and the rotated waveforms $%
h_{+}^{\prime }$ and $h_{\times }^{\prime }$ with $\vartheta =\pi /8$, are
plotted in units of $\mu L_{0}/r$ for $\frak{T}$ varying between $0$ (on the
null surface $t=r$) and $1$ (on the future light cone of the evaporation
point at $t=t_{L}$) for an observer latitude of $\theta =\pi /6$. Note that
the vertical scale is the same on all four plots, so that the amplitudes of
the waveforms can be compared by eye.

Figures 4a to 4d are similar, except that now the observer latitude is $%
\theta =\pi /4$. The vertical scale in Figures 4a to 4d is larger than the
vertical scale in Figures 3a to 3d by a factor of about $1.33$ to $1$.
Figures 5a to 5d are again similar, except with $\theta =\pi /3$. The
vertical scale in Figures 5a to 5d is larger than that in Figures 4a to 4d
by another factor of about $1.13$ to $1$. It is evident that the peak
amplitude of $h_{\times }$ increases monotonically as $\theta $ varies from $%
\pi /6$ to $\pi /3$, but that the peak amplitude of $h_{+}$ remains roughly
constant over this range -- a reflection of the fact that $h_{+}\rightarrow
0 $ at both latitude extremes ($\theta =0$ and $\theta =\pi /2$), in
accordance with (\ref{hcross-asymp}) and (\ref{hcross}).

Figure 6b plots $h_{\times }$ for $\theta =\pi /2$. Because $%
h_{+}\rightarrow 0$ like $\cos \theta $ near the equatorial plane, Figure 6a
plots not $h_{+}$ itself but $\lim_{\theta \rightarrow \pi /2}(h_{+}/\cos
\theta )$. The vertical scale in Figures 6a and 6b is about $1.11$ times
larger than in Figures 5a to 5d, and overall about $1.67$ times larger than
in Figures 3a to 3d. The rotated waveforms $h_{+}^{\prime }$ and $h_{\times
}^{\prime }$ are not shown in the case $\theta =\pi /2$ because they are
just $-\frac{1}{\sqrt{2}}h_{\times }$ and $\frac{1}{\sqrt{2}}h_{\times }$,
respectively.

Apart from the changes in amplitude, the most obvious trend in the shape of
the waveforms is the departure from the simple modulated cosine and sine
waves visible at small values of $\theta $ to the noticeably spiked (in the
case of $h_{\times }$) and bulbous (in the case of $h_{+}/\cos \theta $)
waveforms near $\theta =\pi /2$. Allen and Ottewill \cite{Allen-Ott} have
plotted the linearly polarized waveforms from the stationary ACO loop over a
single cycle for the latitude angles $\theta =\pi /2$, $\pi /4$ and $\pi /20$%
, and their results are recognizably the same as those shown here, although
(of course) Allen and Ottewill did not include the frequency acceleration
induced by the logarithmic dependence of $\psi $ on $t$, or the linear
modulation factor $(r-t+t_{L})/t_{L}$. Oddly, Allen and Ottewill's plot of $%
h_{+}/\sin \theta $ does not vanish at $\theta =\pi /2$, even though they
state explicitly in the text that it should. This discrepancy is presumably
an artefact of their reliance on truncated Fourier series to plot the
waveforms.

For the sake of visual clarity, the parameter $\kappa \mu $ appearing in the
expression for $\psi $ has in Figures 3, 4, 5 and 6 been assigned the value $%
10^{-1}$. This is of course not a realistic value for $\kappa \mu $. Given
that $\kappa \approx 3.\,103\,7$ for the ACO loop, and $\mu $ is expected to
be of order $10^{-6}$ or smaller for a cosmic string formed at a GUT
symmetry-breaking phase transition, the roughly half-dozen cycles visible in
each of the graphs in Figures 3, 4, 5 and 6 should be replaced by at least $%
200,000$ cycles.

\section{Detectability}

The lifetime $t_{L}\equiv \tfrac{1}{4\pi \kappa \mu }L_{0}$ of an
evaporating ACO string loop obviously depends on the values assumed for $\mu 
$ (which with dimensional units restored is $G\mu /c^{2}$) and the loop's
initial length $L_{0}$. If they were ever present in the early Universe, GUT
cosmic strings would have condensed at about $10^{-35}$ seconds after the
Big Bang. It is also thought that, to a first approximation, the string
network will have evolved towards a scaling solution, in which long strings
are straight on scales smaller than a characteristic length comparable to
the horizon size, and so loops -- which form by the intersection or
self-intersection of long strings -- first appear at the horizon scale.

As an example, ACO string loops with $\mu \sim 10^{-6}$ that formed at the
end of the radiation-dominated era -- about 4000 years after the Big Bang --
would have had $L_{0}\sim 4000$ light years and lifetimes of about $10^{8}$
years. With the waveforms executing a total of $200,000$ cycles, this
corresponds to a frequency of about $10^{-9}$ Hz at the time of formation,
although subsequent redshifting of the gravitons by a factor of $z_{\text{eq}%
}^{-1}$ (where $z_{\text{eq}}\approx 1.2\times 10^{4}$ for a universe with
critical density and a Hubble constant of $73$ km/s/Mpc \cite{BCS}) would
reduce this to a currently observed frequency of about $10^{-13}$ Hz. At an
observer distance of $r\sim 10^{10}$ light years, the units $\mu L_{0}/r$
along the vertical scale in Figures 3, 4, 5 and 6 would be $4\times 10^{-13}$%
. Loops formed at earlier epochs and still extant at the end of the
radiation-dominated era would have been smaller, and their frequencies
correspondingly higher and astrophysically more interesting (as millisecond
pulsar timing is sensitive to frequencies of about $10^{-8}$ Hz, while the
LIGO frequency window extends from about $10$ to $10^{4}$ Hz), but their
amplitudes would have been proportionally smaller, given that $h_{+}$ and $%
h_{\times }$ scale as $L(t)$.

However, it is not the waveform amplitudes themselves that determine the
detectability of a source of gravitational radiation, but rather the density
of gravitational energy $\rho _{\text{g}}$ per logarithmic frequency
interval, or spectral density, which is conventionally normalized as $\Omega
_{\text{g}}(f)=\rho _{\text{c}}^{-1}(fd\rho _{\text{g}}/df)$, where $\rho _{%
\text{c}}$ is the critical energy density of the Universe. An upper bound on
the magnitude of $\Omega _{\text{g}}$ for a single string loop can be
estimated as follows.

A loop forming at a time $t_{\text{F}}$ after the Big Bang would have an
initial length $L_{0}\sim ct_{\text{F}}$. If the loop radiates a power $P$
over a small time interval $\Delta t$ centered on a time $t_{\text{R}}\geq
t_{\text{F}}$, then at the current time $t_{0}\gg t_{\text{R}}$ the total
gravitational energy output $P\Delta t$ would be distributed over a
spherical shell with volume $\Delta V=4\pi c^{3}a(t_{0})^{3}a(t_{\text{R}%
})^{-1}[\int_{t_{\text{R}}}^{t_{0}}a(t)^{-1}dt]^{2}\Delta t$, assuming a
spatially-flat Robertson-Walker spacetime with scale factor $a(t)$. The
corresponding mean density of the shell is $\rho _{\text{g}}=P\Delta
t/\Delta V$, and so if $a$ is assumed to have a power-law dependence $%
a(t)\varpropto t^{k}$ with any index $k\in [\frac{1}{2},\frac{2}{3}]$ an
order-of-magnitude estimate of $d\rho _{\text{g}}/dt_{\text{R}}$ turns out
to be: 
\begin{equation}
d\rho _{\text{g}}/dt_{\text{R}}\sim c^{-3}P\,z_{\text{R}}^{-1}\,t_{\text{R}%
}^{-1}t_{0}^{-2}
\end{equation}
(on making the replacement $a(t_{0})/a(t_{\text{R}})\approx z_{\text{R}}$).

Furthermore, if $\gamma ^{0}$ is the radiative efficiency of the loop then
its length at time $t_{\text{R}}$ is $L(t_{\text{R}})\sim ct_{\text{F}%
}-\gamma ^{0}(G\mu /c)(t_{\text{R}}-t_{\text{F}})$, and the corresponding
frequency at the time of emission is $f_{\text{R}}\sim 2c/L(t_{\text{R}})$.
The currently observed emission frequency is therefore $f=a(t_{\text{R}})f_{%
\text{R}}/a(t_{0})$. Since $a(t_{\text{R}})\varpropto t_{\text{R}}^{1/2}$
for emission times during the radiation-dominated era, this leads to the
estimate 
\begin{equation}
f^{-1}df/dt_{\text{R}}\sim t_{\text{R}}^{-1}[\tfrac{1}{2}+\gamma ^{0}(G\mu
/c)t_{\text{R}}/L(t_{\text{R}})]\text{,}
\end{equation}
where the term in square brackets is bounded below by $\tfrac{1}{2}+\gamma
^{0}(G\mu /c^{2})\approx \tfrac{1}{2}$ (this bound being achieved when $t_{%
\text{R}}=t_{\text{F}}$).

So 
\begin{equation}
fd\rho _{\text{g}}/df=(f^{-1}df/dt_{\text{R}})^{-1}(d\rho _{\text{g}}/dt_{%
\text{R}})\lesssim 2c^{-3}P\,z_{\text{R}}^{-1}\,t_{0}^{-2}\text{,}
\label{bound1}
\end{equation}
where (with units restored) $P=\gamma ^{0}(G\mu /c^{2})^{2}c^{5}/G$. Also, a
useful formula relating the redshift factor $z$ at time $t_{\text{R}}$ to
the current density $\rho _{\text{rad}}$ of thermal radiation is \cite
{Vach-Vil}: 
\begin{equation}
z_{\text{R}}^{4}\rho _{\text{rad}}=\tfrac{3c^{2}}{32\pi G}t_{\text{R}}^{-2}%
\text{,}
\end{equation}
which when substituted into (\ref{bound1}) gives: 
\begin{equation}
\Omega _{\text{g}}(f)\equiv \rho _{\text{c}}^{-1}(fd\rho _{\text{g}%
}/df)\lesssim \tfrac{64\pi }{3}\gamma ^{0}(G\mu /c^{2})^{2}\,\Omega _{\text{%
rad}}\,z_{\text{R}}^{3}(t_{\text{R}}/t_{0})^{2}\text{,}
\end{equation}
where $\Omega _{\text{rad}}=\rho _{\text{rad}}/\rho _{\text{c}}$, a
parameter whose value is currently estimated to be $4.6\times 10^{-5}$ \cite
{BCS}.

In particular, if $t_{\text{R}}\approx 4000$ yr -- the time of
radiation/matter pressure equilibrium -- then with $z_{\text{eq}}\approx
1.2\times 10^{4}$ and $t_{0}\approx 1.3\times 10^{10}$ yr, the value of the
spectral density $\Omega _{\text{g}}$ for an ACO loop with $\gamma
^{0}\approx 40$ and $G\mu /c^{2}\approx 10^{-6}$ is bounded above by $%
2\times 10^{-14}$, and scales as $\mu ^{2}$. By way of comparison, the
spectral density of the entire stochastic cosmic string background is
estimated for $G\mu /c^{2}\approx 10^{-6}$ to have the lower bound $\Omega _{%
\text{g}}\gtrsim 1.4\times 10^{-9}$, independently of the observed frequency 
$f$ for a wide range of frequencies \cite{CBS}. The space-based
interferometer LISA is projected to have a peak sensitivity of $\Omega _{%
\text{g}}\approx 10^{-11}$ at a frequency of $10^{-3}$ Hz \cite{DeP-Hog},
while the estimated strain sensitivity of Advanced LIGO (due to go on-line
in 2013) at frequencies around $100$ Hz corresponds to $\Omega _{\text{g}%
}\approx 3\times 10^{-10}$ \cite{CBS}

There is of course nothing peculiar about either the frequencies or
amplitudes of the waveforms emitted by evaporating ACO\ loops, save that
their lifetimes are longer, their waveform amplitudes are typically smaller,
and their spectra presumably more stable than those of other loop
configurations with the same initial length. More detailed and realistic
estimates of the effects and detectability of gravitational radiation from
them can therefore be inferred from similar studies in standard references
(for example, \cite{Vach-Vil, All-Casp, CBS, BCS, DeP-Hog}). The current
consensus is that the strain sensitivity of LIGO is currently one or two
orders of magnitude too large to detect any stochastic background from
cosmic string loops with $G\mu /c^{2}\gtrsim 10^{-6}$, but that millisecond
pulsar spindown rates have already ruled out relic strings with $G\mu
/c^{2}\gtrsim 10^{-9}$. It is therefore unlikely that individual string
spectra will ever be visible to current or foreseeable gravitational wave
observatories.

\section{Conclusions}

As has already been mentioned in \cite{Anderson5}, the ACO loop is one of
the most important flat-space cosmic string solutions, as it is possibly the
longest-lived of all loop configurations, and is moreover the only loop
solution (bar the collapsing circular loop) whose evolution is known to be
analytically tractable. The fact that the linearly polarized waveforms $%
h_{+} $ and $h_{\times }$ emitted by the evaporating ACO loop can be
represented by the simple exact forms (\ref{hplus}) and (\ref{hcross}) is an
added bonus.

The actual shapes of the waveforms, as plotted in Figures 3, 4, 5 and 6, are
admittedly not very surprising, as Allen and Ottewill have already
calculated and plotted the linearly polarized waveforms from the stationary
ACO loop in \cite{Allen-Ott}. However, their formulas for $h_{+}$ and $%
h_{\times }$ involved Fourier series rather than closed-form expressions and
did not include the effects of frequency acceleration or amplitude
modulation. What has been presented here is therefore a more complete
treatment of the problem for this particular, and very significant, loop
configuration.

LIST\ OF\ FIGURE CAPTIONS:\bigskip

Figure 1: $y$-$z$ projection of the Allen-Casper-Ottewill loop at times $%
\tau -\varepsilon =0$,

$L/16$, $L/8$ and $3L/16$ (top row) and $\tau -\varepsilon =L/4$, $5L/16$, $%
3L/8$ and $7L/16$

(bottom row), where the time offset $\varepsilon $ is $0.02L$. The string
has been artificially

thickened for the sake of visibility, and the $z$-axis is also shown.\bigskip

Figure 2: Schematic representation of the evaporation of the ACO loop. The

thickened line corresponds to the outer envelope of the loop, and $F_{0}$
and $F_{L}$

are the future light cones of the origin and the evaporation point
respectively.\bigskip

Figure 3a: The waveform $h_{+}$ (in units of $\mu L_{0}/r$) plotted against $%
\frak{T}$ for $\theta =\pi /6$

and $\vartheta =0$.\bigskip

Figure 3b: The waveform $h_{\times }$ (in units of $\mu L_{0}/r$) plotted
against $\frak{T}$ for $\theta =\pi /6$

and $\vartheta =0$.\bigskip

Figure 3c: The waveform $h_{+}^{\prime }$ (in units of $\mu L_{0}/r$)
plotted against $\frak{T}$ for $\theta =\pi /6$

and $\vartheta =\pi /8$.\bigskip

Figure 3d: The waveform $h_{\times }^{\prime }$ (in units of $\mu L_{0}/r$)
plotted against $\frak{T}$ for $\theta =\pi /6$

and $\vartheta =\pi /8$.\bigskip

Figure 4a: The waveform $h_{+}$ (in units of $\mu L_{0}/r$) plotted against $%
\frak{T}$ for $\theta =\pi /4$

and $\vartheta =0$.\bigskip

Figure 4b: The waveform $h_{\times }$ (in units of $\mu L_{0}/r$) plotted
against $\frak{T}$ for $\theta =\pi /4$

and $\vartheta =0$.\bigskip

Figure 4c: The waveform $h_{+}^{\prime }$ (in units of $\mu L_{0}/r$)
plotted against $\frak{T}$ for $\theta =\pi /4$

and $\vartheta =\pi /8$.\bigskip

Figure 4d: The waveform $h_{\times }^{\prime }$ (in units of $\mu L_{0}/r$)
plotted against $\frak{T}$ for $\theta =\pi /4$

and $\vartheta =\pi /8$.\bigskip

Figure 5a: The waveform $h_{+}$ (in units of $\mu L_{0}/r$) plotted against $%
\frak{T}$ for $\theta =\pi /3$

and $\vartheta =0$.\bigskip

Figure 5b: The waveform $h_{\times }$ (in units of $\mu L_{0}/r$) plotted
against $\frak{T}$ for $\theta =\pi /3$

and $\vartheta =0$.\bigskip

Figure 5c: The waveform $h_{+}^{\prime }$ (in units of $\mu L_{0}/r$)
plotted against $\frak{T}$ for $\theta =\pi /3$

and $\vartheta =\pi /8$.\bigskip

Figure 5d: The waveform $h_{\times }^{\prime }$ (in units of $\mu L_{0}/r$)
plotted against $\frak{T}$ for $\theta =\pi /3$

and $\vartheta =\pi /8$.\bigskip

Figure 6a: The rescaled waveform $h_{+}/\cos \theta $ (in units of $\mu
L_{0}/r$) in the limit as

$\theta \rightarrow \pi /2$ plotted against $\frak{T}$ for $\vartheta =0$%
.\bigskip

Figure 6b: The waveform $h_{\times }$ (in units of $\mu L_{0}/r$) plotted
against $\frak{T}$ for $\theta =\pi /2$

and $\vartheta =0$.\bigskip

\newpage

\begin{figure}
\epsfig{file=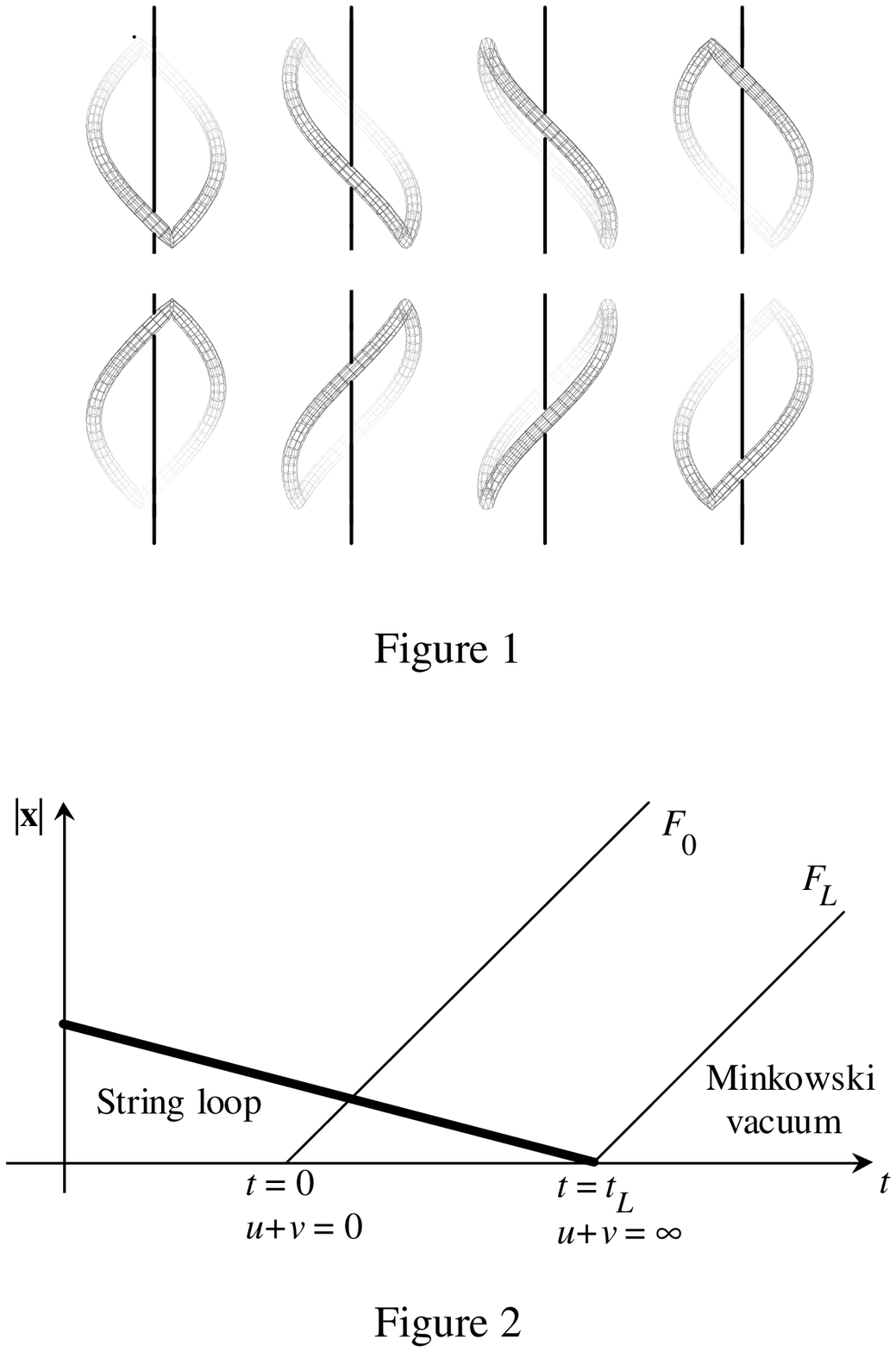, width=160mm}
\end{figure}

\newpage

\begin{figure}
\epsfig{file=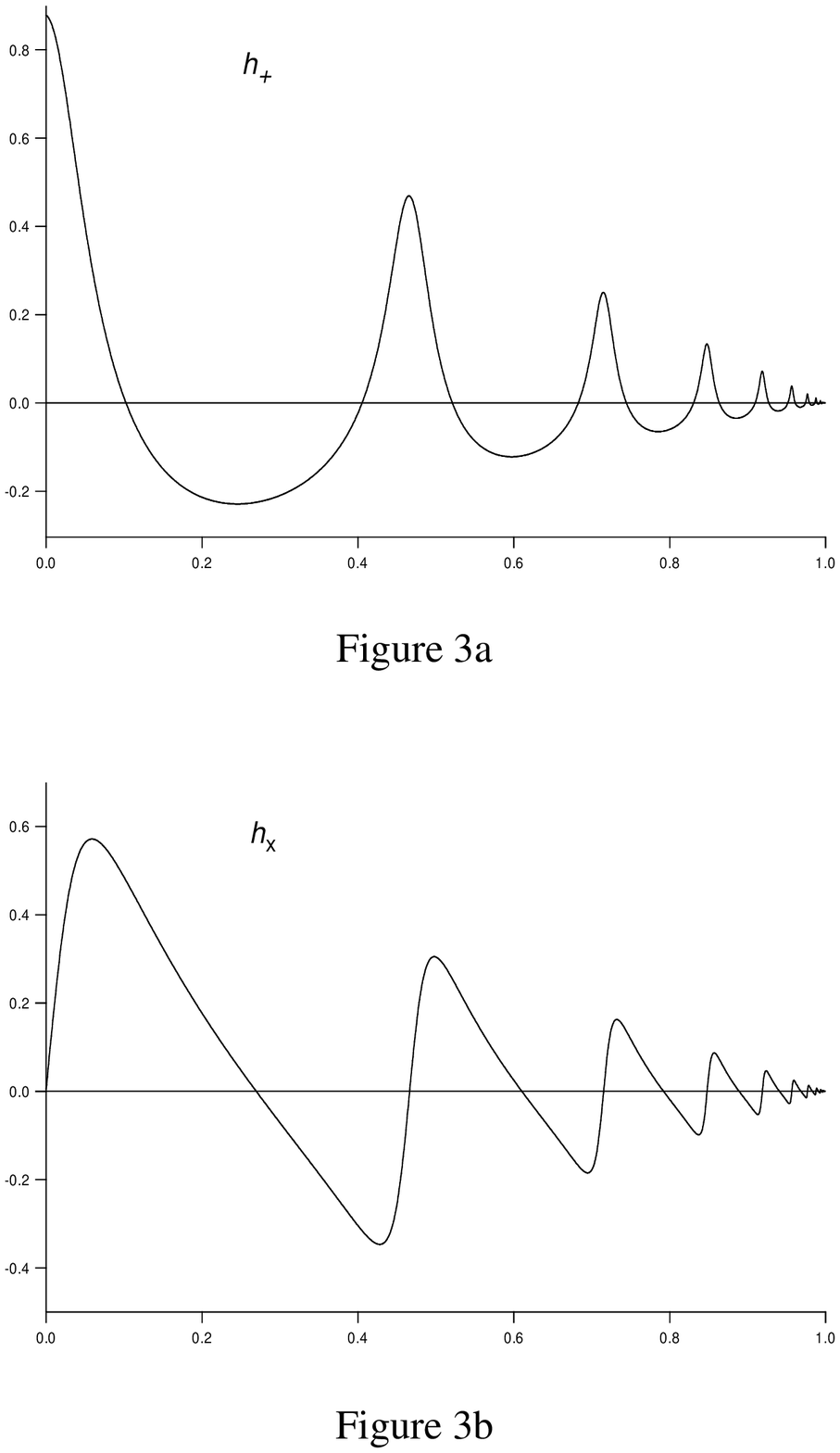, width=160mm}
\end{figure}

\newpage

\begin{figure}
\epsfig{file=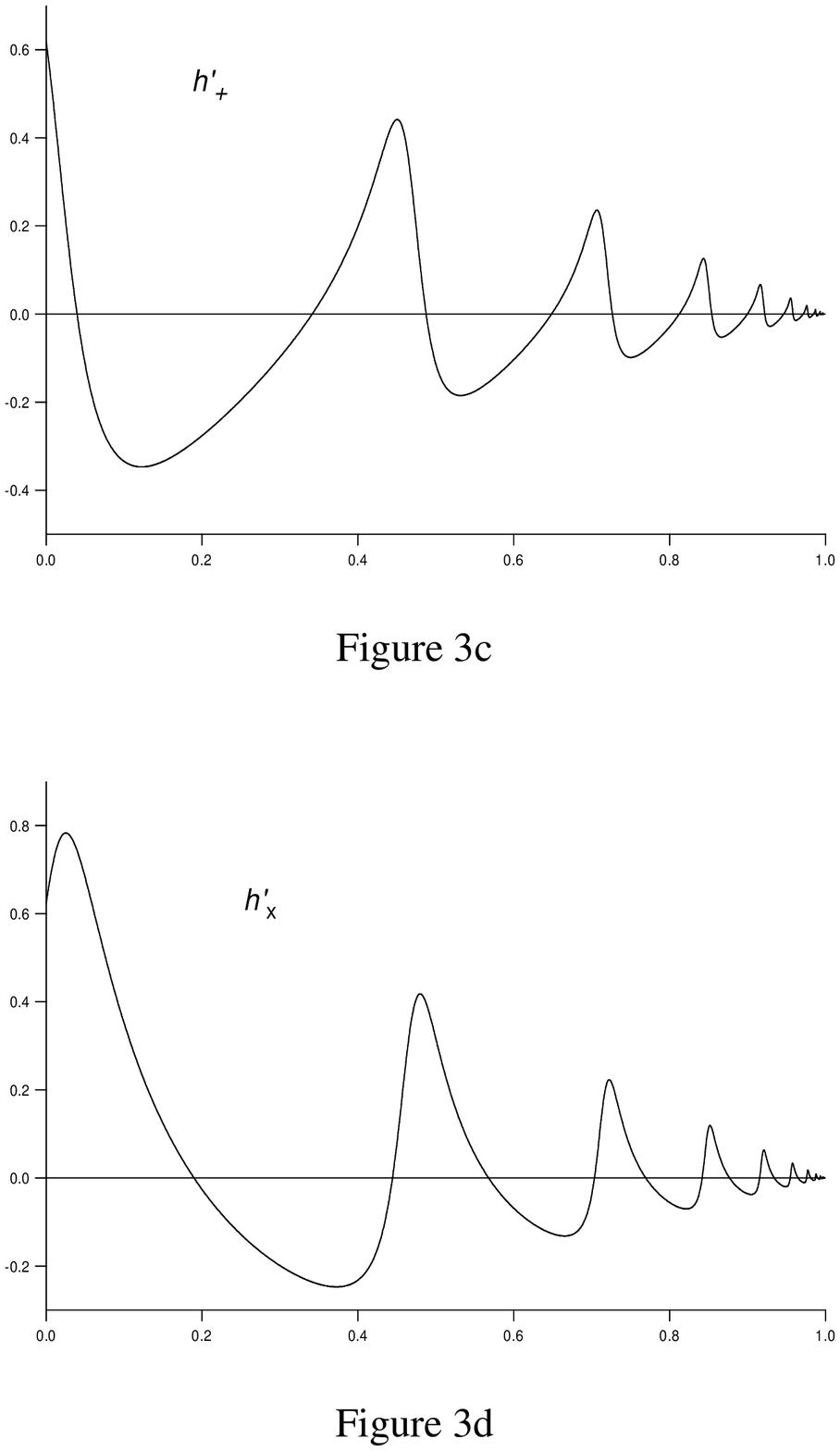, width=160mm}
\end{figure}

\newpage

\begin{figure}
\epsfig{file=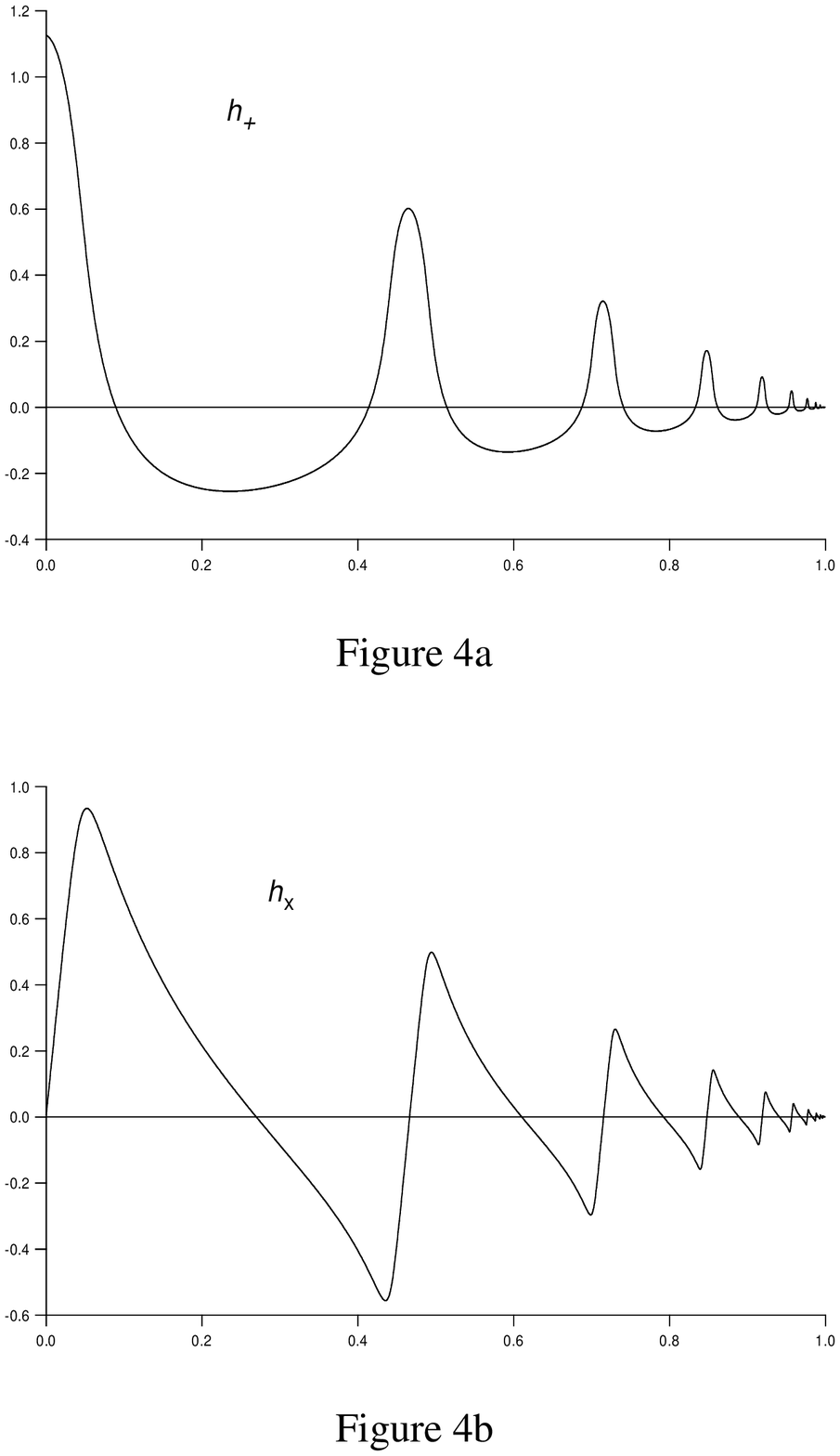, width=160mm}
\end{figure}

\newpage

\begin{figure}
\epsfig{file=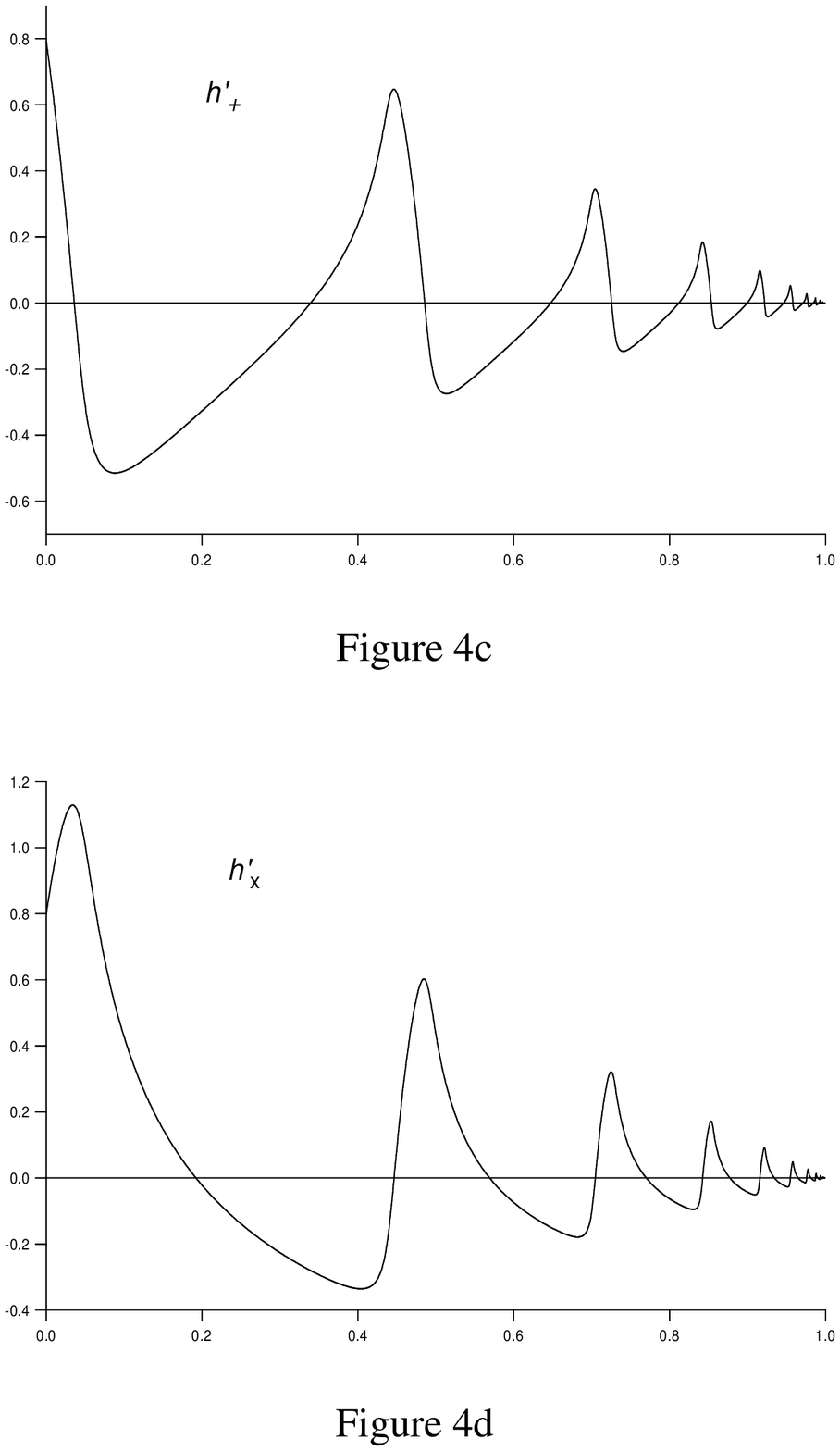, width=160mm}
\end{figure}

\newpage

\begin{figure}
\epsfig{file=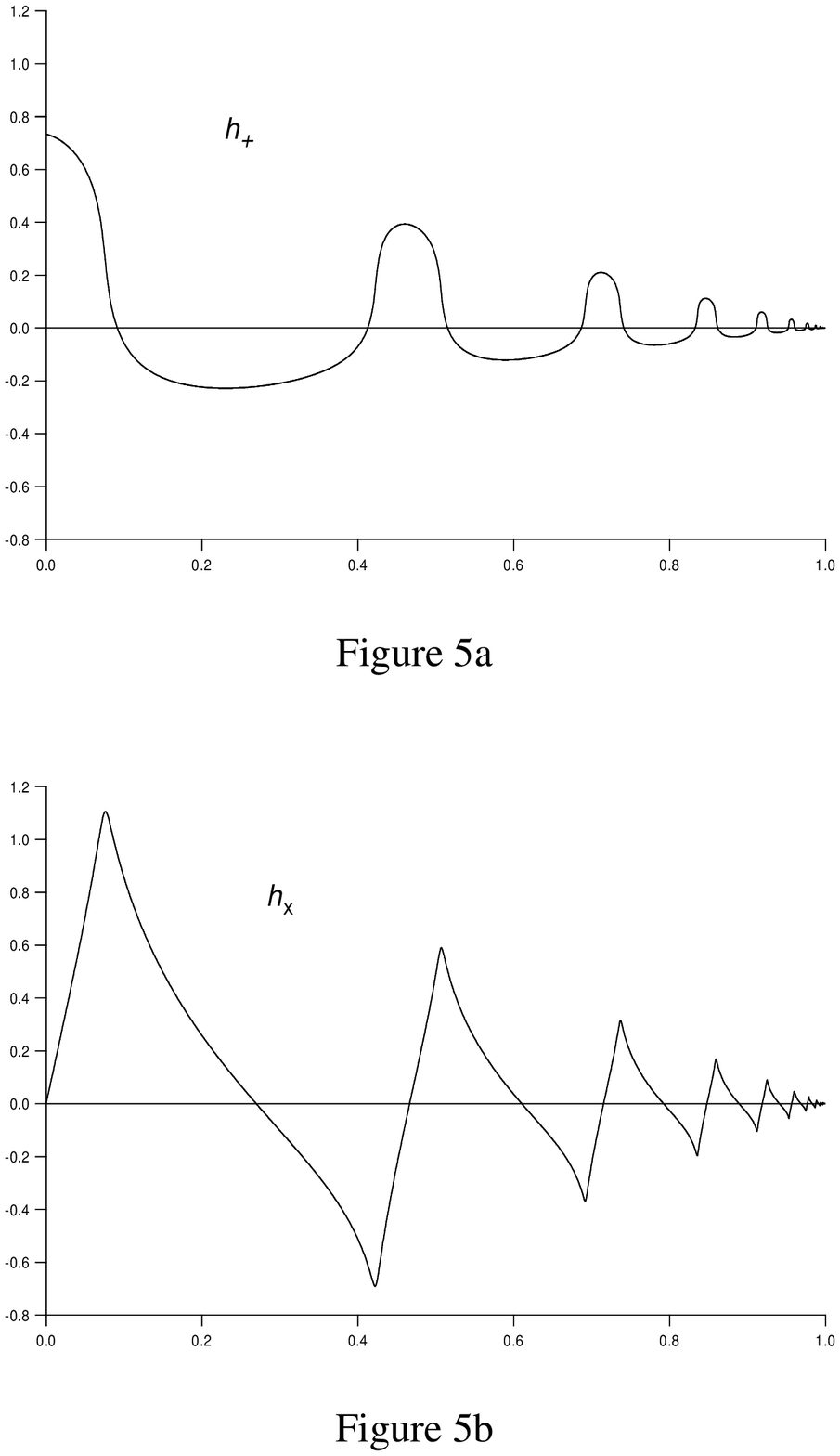, width=160mm}
\end{figure}

\newpage

\begin{figure}
\epsfig{file=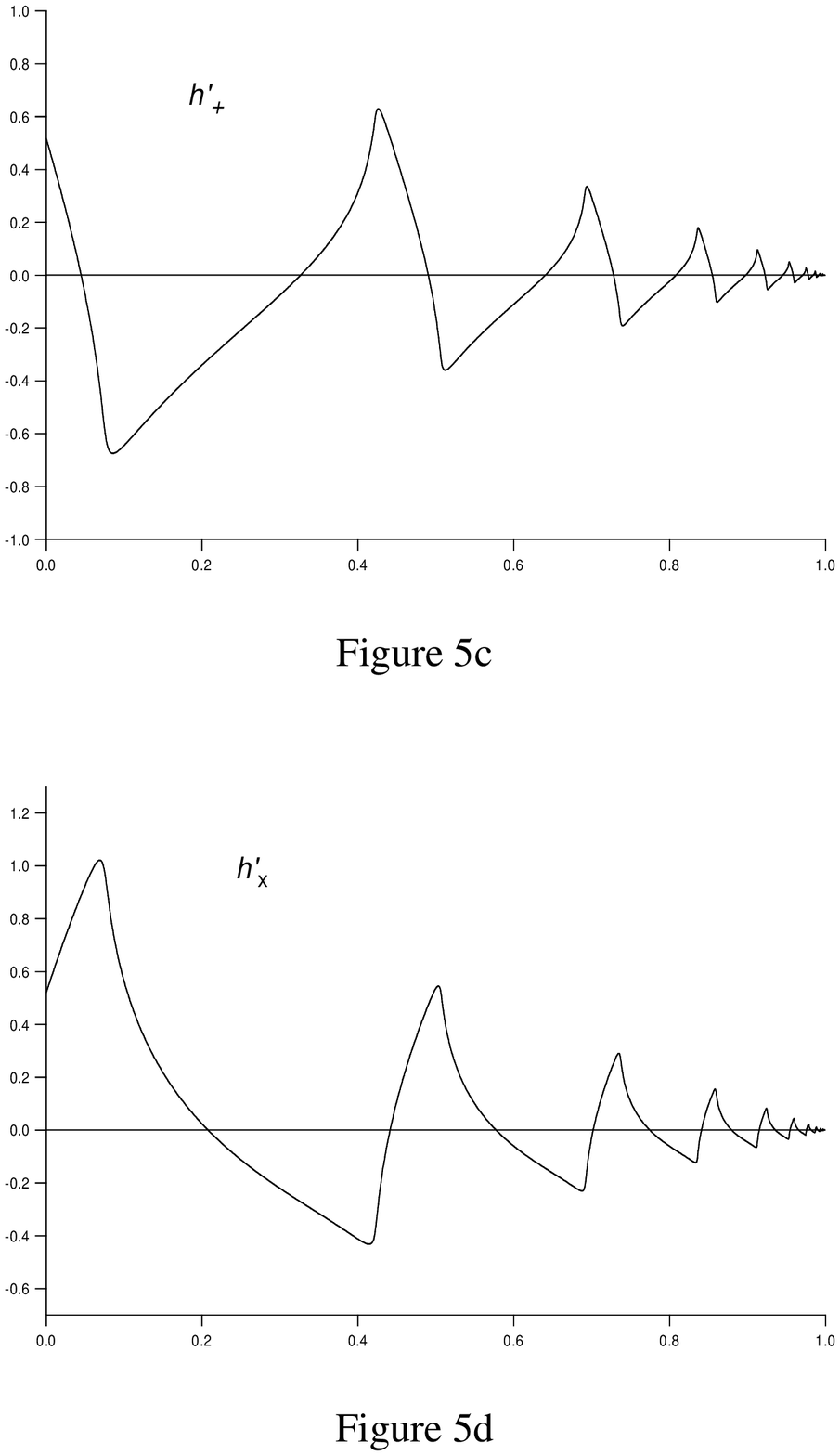, width=160mm}
\end{figure}

\newpage

\begin{figure}
\epsfig{file=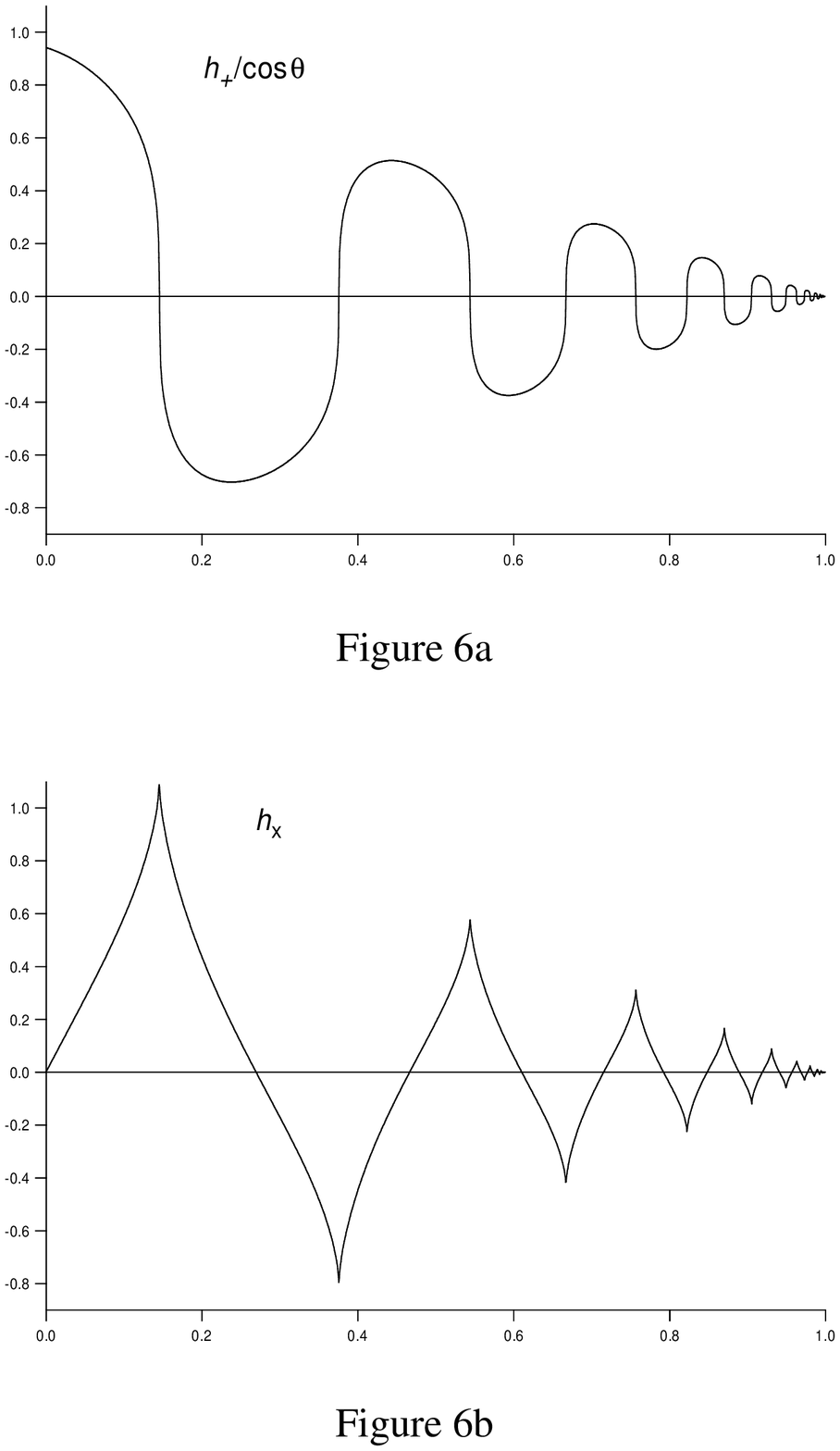, width=160mm}
\end{figure}

\end{document}